\documentclass[aps,prd,prev,twocolumn,superscriptaddress,preprintnumbers,floatfix,nofootinbib]{revtex4-1}
\pdfoutput=1
\usepackage{graphicx}
\usepackage{bm}
\usepackage{times}
\usepackage{hyperref}
\usepackage{slashed}
\usepackage{color}
\usepackage{aas_macros}
\usepackage{slashed}
\usepackage{lipsum}
\usepackage{subfigure}
\usepackage{multirow}
\usepackage{amsmath}
\usepackage{array} % for defining a new column type
\usepackage{varwidth} %for the varwidth minipage environment

\newcommand{\be}{\begin{equation}}
\newcommand{\ee}{\end{equation}}
\newcommand{\bea}{\begin{eqnarray}}
\newcommand{\eea}{\end{eqnarray}}

\begin{document}

\hspace{13cm} \parbox{5cm}{FERMILAB-PUB-17-080-A}

\hspace{13cm}
\vspace{0.2cm}

%\title{A Single, Steady-State Model for the TeV Gamma-Ray Emission from the Galactic Center}
\title{Using HAWC to Discover Invisible Pulsars}
\author{Tim Linden}
\email{linden.70@osu.edu, orcid.org/0000-0001-9888-0971}
\affiliation{Center for Cosmology and AstroParticle Physics (CCAPP), and \\ Department of Physics, The Ohio State University Columbus, OH, 43210}
\author{Katie Auchettl}
\email{auchettl.1@osu.edu, orcid.org/0000-0002-4449-9152}
\affiliation{Center for Cosmology and AstroParticle Physics (CCAPP), and \\ Department of Physics, The Ohio State University Columbus, OH, 43210}
\author{Joseph Bramante}
\email{jbramante@perimeterinstitute.ca, orcid.org/0000-0001-8905-1960}
\affiliation{Perimeter Institute for Theoretical Physics, Waterloo, Ontario, N2L 2Y5, Canada}
\author{Ilias Cholis}
\email{icholis1@jhu.edu, orcid.org/0000-0002-3805-6478}
\affiliation{Department of Physics and Astronomy, The Johns Hopkins University, Baltimore, Maryland, 21218}
\author{Ke~Fang}
\email{kefang@umd.edu, orcid.org/0000-0002-5387-8138}
\affiliation{University of Maryland, Department of Astronomy, College Park, MD, 20742}
\affiliation{Joint Space-Science Institute, College Park, MD, 20742} 
\author{Dan Hooper}
\email{dhooper@fnal.gov, orcid.org/0000-0001-8837-4127}
\affiliation{Fermi National Accelerator Laboratory, Center for Particle Astrophysics, Batavia, IL 60510}
\affiliation{University of Chicago, Department of Astronomy and Astrophysics, Chicago, IL 60637}
\affiliation{University of Chicago, Kavli Institute for Cosmological Physics, Chicago, IL 60637}
\author{Tanvi Karwal} 
\email{tkarwal@jhu.edu, orcid.org/0000-0002-1384-9949}
\affiliation{Department of Physics and Astronomy, The Johns Hopkins University, Baltimore, Maryland, 21218}
\author{Shirley Weishi Li}
\email{li.1287@osu.edu, orcid.org/0000-0002-2157-8982}
\affiliation{Center for Cosmology and AstroParticle Physics (CCAPP), and \\ Department of Physics, The Ohio State University Columbus, OH, 43210}

\begin{abstract}
Observations by HAWC and Milagro have detected bright and spatially extended TeV $\gamma$-ray sources surrounding the Geminga and Monogem pulsars. We argue that these observations, along with a substantial population of other extended TeV sources coincident with pulsar wind nebulae, constitute a new morphological class of spatially extended TeV halos. We show that HAWCs wide field-of-view unlocks an expansive parameter space of TeV halos not observable by atmospheric Cherenkov telescopes. Under the assumption that Geminga and Monogem are typical middle-aged pulsars, we show that ten-year HAWC observations should eventually observe 37$^{+17}_{-13}$ middle-aged TeV halos that correspond to pulsars whose radio emission is not beamed towards Earth. Depending on the extrapolation of the TeV halo efficiency to young pulsars, HAWC could detect more than 100 TeV halos from mis-aligned pulsars. These pulsars have historically been difficult to detect with existing multiwavelength observations.  TeV halos will constitute a significant fraction of all HAWC sources, allowing follow-up observations to efficiently find pulsar wind nebulae and thermal pulsar emission. The observation and subsequent multi-wavelength follow-up of TeV halos will have significant implications for our understanding of pulsar beam geometries, the evolution of PWN, the diffusion of cosmic-rays near energetic pulsars, and the contribution of pulsars to the cosmic-ray positron excess.
\end{abstract}

\maketitle

\section{Introduction}
%We have detected a class of TeV nebulae
Recent observations by the High Altitude Water Cherenkov Observatory (HAWC)~\cite{Abeysekara:2017hyn}, along with earlier results from Milagro~\cite{2009ApJ...700L.127A}, have detected diffuse TeV emission surrounding the Geminga and B0656+14 (hereafter referred to as Monogem~\citep{2003ApJ...592L..71T}) pulsars. While it is difficult to constrain the exact morphology of this emission, both systems are well-fit by Gaussian distributions with an angular extension of $\sim$2$^\circ$. These observations are intriguing for several reasons. First, the short cooling times of very high energy electrons imply that even middle-aged pulsars accelerate e$^+$e$^-$ to energies exceeding $\sim$50~TeV. Second, the angular size of these ``TeV halos" indicates that the propagation of cosmic rays near pulsars is significantly more constrained than typical for the interstellar medium~\citep{2017arXiv170208436H, hawctbs}. Third, the intensity of this emission indicates that a significant fraction of the total pulsar spin-down luminosity is converted into e$^+$e$^-$ pairs, providing evidence in support of pulsar interpretations of the rising cosmic-ray positron fraction observed by PAMELA and \mbox{AMS-02}~\citep{Adriani:2008zr, Aguilar:2013qda, Yuksel:2008rf, 2017arXiv170208436H}. 

%%TeV PWN Observations
The observation of  extended ``TeV halos" surrounding Geminga and Monogem augment a growing class of TeV sources coincident with pulsars and pulsar wind nebulae (PWN). To date, Atmospheric Cherenkov Telescopes (ACTs), such as H.E.S.S. and VERITAS, have discovered a population of at least 32 such sources~\citep{2006csxs.book..279K,  2008ICRC....2..659C, 2009arXiv0912.4304H, 2013arXiv1305.2552K}\footnote{http://tevcat.uchicago.edu/}. H.E.S.S. refers to these sources as ``TeV PWN", noting that the TeV emission is correlated with pulsars that have visible PWN. However, results from the H.E.S.S. collaboration indicate that the TeV emission is significantly more extended than the X-ray PWN~\citep{Abdalla:2017vci}. Thus, these systems may have a unique origin, morphology, and dynamical evolution.

H.E.S.S. observations indicate two important features of TeV halos. First, there is a close correlation between the pulsar spin-down luminosity and the luminosity of the TeV halo. Second, the physical size of the TeV halo is correlated to the pulsar age~\citep{Abdalla:2017vci}. While ACT observations have been extremely efficient in finding TeV features surrounding known radio pulsars with high spin-down luminosities ($\dot{E}$~$>$~10$^{36}$~erg~s$^{-1}$), they have struggled to find nearby pulsars with lower spindown luminosities. The H.E.S.S. sensitivity to TeV halos degrades significantly for systems with angular extensions exceeding 0.6$^\circ$~\citep{Abdalla:2017vci}. Indeed, the 2$^\circ$ TeV halos surrounding Geminga and Monogem have not been detected by ACT instrumentation despite their high flux. 

%Undetected sources. 
Intriguingly, observations of TeV halos provide a new avenue to discover pulsars. To date, the vast majority of pulsars have been discovered based on their beamed radio emission. However, these systems are only the tip of the iceberg; a substantial population of ``invisible" pulsars with mis-aligned beaming angles lurks below. Mis-aligned pulsars have traditionally been extremely difficult to detect. In particular, the effectiveness of ACTs in finding mis-aligned pulsars is inhibited by their small field-of-view. This suggests that HAWC observations can reveal an untapped parameter space, finding a significant population of nearby pulsars via their spatially extended TeV halos.

In this paper, we first argue that TeV halos are a generic feature of pulsars. Second, we show that current searches for mis-aligned pulsars suffer from significant incompleteness, even for nearby systems. Combining these two results, we conclude that HAWC is likely to detect many unassociated TeV halos, with a particular advantage in detecting middle-aged\footnote{Throughout the paper we define middle-aged pulsars to be those with characteristic ages between 100-400~kyr.} pulsars in close proximity to Earth. We discuss multi-wavelength observations capable of confirming the nature of these sources. HAWC's wide field-of-view will provide sensitivity to TeV halos over nearly half the sky, providing new insights into the size of pulsar radio and $\gamma$-ray beams, the evolution of PWN, the cosmic-ray diffusion parameters near compact objects, and the contribution of pulsars to the cosmic-ray positron excess.

%%Paper outline
The paper is outlined as follows. In Section~\ref{sec:tevhalos} we establish that TeV halos constitute a new morphological class of TeV emission sources, with characteristics that are distinct from both supernova remnants (SNR) and PWN. In Sections~\ref{sec:acthalos}~and~\ref{sec:hawcsensitivity} we discuss the sensitivity of H.E.S.S and HAWC to TeV halos, respectively. Our analysis indicates that HAWC is sensitive to an expansive population of TeV halos not observable by ACTs. In Section~\ref{sec:sourcenumbers} we discuss the population of ``invisible" pulsars which are not detected because their radio beams are not oriented towards Earth. Utilizing the observation of TeV halos coincident with previously known pulsars, we argue that TeV halo observations will uncover a large population of these systems. In Section~\ref{sec:pwn} we discuss X-Ray and optical observations that could definitively prove the TeV halo origin of these hidden sources. In Sections~\ref{sec:youngpwn}~and~\ref{sec:discussion} we note that the unbiased nature of TeV halo observations provides new insight into multiple open questions concerning the evolution of pulsars, the diffusion parameters of the Milky Way, and the origin of the cosmic-ray positron excess.

\section{TeV Halos are a New Morphological Feature}
\label{sec:tevhalos}

In this section, we argue that the population of extended TeV sources coincident with known pulsars constitutes a new morphological source class. We name these sources TeV halos. We note that the radial extent of TeV halos is significantly smaller than their associated SNR~\citep{1996ApJ...463..224P,1998ApJ...500..342B}, but significantly larger than PWN~\citep{2006ARA&A..44...17G}. Observations of TeV halos corresponding to middle-aged pulsars are crucial to differentiate these sources, as the morphological differences between each source class become more pronounced over time.

The morphology of supernova remnants is determined by the interactions of the supernova-powered shock front with the interstellar medium. The size of supernova remnants expands monotonically, though the rate varies as the density of the interstellar medium and radiative losses inside the remnant begin to play more important roles~\citep[see e.g.][]{2012A&ARv..20...49V}. For middle-aged SNR near the solar position, a radius $\gtrsim$50~pc is typical. This significantly exceeds the $\sim$10~pc extensions observed in middle-aged TeV halos. This distinction is particularly stark for Monogem; the SNR is found to be extended by $\sim$25$^\circ$~\citep{1981ApJ...248..152N}, while the TeV halo is $\sim$2$^\circ$.  In addition to their different sizes, TeV halos are found to be offset from the centers of SNR. This is expected, as pulsars are born with typical kick velocities of $\sim$400~km~s$^{-1}$. For ages of $\sim$100~kyr, this translates to average offsets of $\sim$40~pc. Observations of TeV halos by high-resolution ACTs indicate that the emission is more closely correlated to the pulsar than the SNR~\cite{Abdalla:2017vci}.

The radial extent of PWN is similarly determined by the interactions of the pulsar's relativistic wind with the surrounding medium. In the early stages of pulsar evolution, the pulsar is confined within the SNR, where it expands freely for up to $\sim$10$^4$~yr until the reverse SNR shock compresses it. These interactions, along with the significant kick velocity of the pulsar make the morphology of early PWN extremely complex. Simulations indicate that the PWN can reach radial extents up to $\sim$4~pc~\cite{Chevalier:1977wd, 1984ApJ...278..630R, vanderSwaluw:2000ei, Blondin:2001bf, vanderSwaluw:2003gg}. However, pulsars with kick velocities \mbox{$v \gtrsim 100$~km/s} and \mbox{ages $\gtrsim 10^{5}$~yr} have typically exited the SNR \cite{vanderSwaluw:2002ah} and  have begun interacting with the diffuse interstellar medium. Thus, their termination shock front, which is similar to their forward shock, is constrained by the ram pressure of the interstellar medium. The radius of the PWN can be calculated as:

%PWN can be calculated in two stages. The first occurs while a pulsar is confined within a supernova remnant in the Sedov-Taylor stage, and is defined by a PWN radius that increases quickly as either (R~$\propto$~t$^{11-15}$) or 

%quickly (R~$\propto$~t$^{6/5}$) into the low-density SNR. Numerical simulations indicate that PWN can reach radial extents of $\sim$4~pc at ages of $\sim$30~kyr~\citep{vanderSwaluw:2000ei}. However, middle-aged pulsars are expected to 

%However, pulsars with kick velocities \mbox{$v \gtrsim 100$~km/s} with \mbox{ages $\gtrsim 10^{5}$~yr} are expected to have exited the SNR shock front \cite{vanderSwaluw:2002ah} and to have begun interacting with the diffuse interstellar medium. Thus, their termination shock front, which is similar to their forward shock, is constrained by the ram pressure of the interstellar medium. The radius of the PWN can be calculated as:

\begin{multline}
\label{eq:pwnradius}
R_{\textrm{PWN}} \simeq  1.5 \left( \frac{\dot{E}}{10^{35} \, \textrm{erg/s}} \right)^{1/2} \times \\  \left( \frac{n_{\textrm{gas}}}{1 \, \textrm{cm}^{-3}} \right)^{-1/2}  \left( \frac{v}{100 \, \textrm{km/s}} \right)^{-3/2}  \textrm{pc}
\end{multline}

\noindent where $\dot{E}$ is the spin-down power of the pulsar, n$_{gas}$ is the local gas density in the interstellar medium, and $v$ is the pulsar kick velocity. This falls below the radial extent of middle-aged TeV halos by approximately one order of magnitude. In the case of Geminga, this difference is striking; the PWN is observed to be confined within $\sim$2~arcminutes of the central pulsar~\citep{Posselt:2016lot}, while the TeV halo extends for $\sim$2$^\circ$.

Thus, standard models of SNR and PWN evolution do not explain the existence of a morphological feature extending for $\sim$10~pc and centered near the pulsar location. ACT observations indicate that the dynamics of this region are complex. For example, TeV halos are found to be centered at a position offset from their associated pulsar. The degree of this offset is found to increase with the pulsar age~\cite{Abdalla:2017vci}. It is clear that the SNR, PWN, and interstellar medium all play important roles in determining the extent, morphology, and characteristics of TeV halos, and further investigations are needed to understand the dynamics of this region.

While the physical source of the TeV halo region is unknown, observations indicate that cosmic-ray diffusion inside TeV halos is significantly inhibited compared to the surrounding interstellar medium. In particular, the luminosity of the TeV halo associated with Geminga implies that $\sim$7-29\% of the pulsar spin-down energy is converted into e$^+$e$^-$ pairs in order to generate the $\gamma$-ray signal~\cite{2017arXiv170208436H}. If a significant fraction of the e$^+$e$^-$ energy escapes from the TeV halo, the necessary e$^+$e$^-$ injection power would exceed the total pulsar spin-down energy. Thus, e$^+$e$^-$ must remain confined within TeV halos for a significant fraction of an energy loss time. 

Given that the pulsar spin-down power is incapable of providing significantly enhanced magnetic field or interstellar radiation field (ISRF) energy densities on $\sim$10~pc scales~\cite{2017arXiv170208436H}, we adopt typical values for the magnetic field and interstellar radiation field (ISRF) energy densities near the solar position ($\rho_{{\rm mag}}$~=~0.224~eV~cm$^{-3}$, $\rho_{{\rm ISRF}}$~=~1.56~eV~cm$^{-3}$~\citep{Porter:2005qx}). The energy loss time of $\sim$10~TeV particles is then (temporarily ignoring $\mathcal{O}(1)$ effects such as Klein-Nishina suppression of inverse-Compton scattering):

\begin{equation}
\label{eq:losstime}
\tau_{{\rm loss}}  \approx 2~\times~10^4~{\rm yr}~\left (\frac{10~{\rm TeV}}{E_e} \right)
\end{equation}

\noindent We immediately note two implications. First, requiring that particles diffuse only $\sim$10~pc in 2$\times$10$^{4}$~yr implies that the diffusion coefficient at 10~TeV that is no larger than 2.5$\times$10$^{26}$~cm$^{2}$s$^{-1}$. Compared to standard diffusion parameters in the interstellar medium (e.g. D$_0$~$\approx$~5$\times$10$^{28}$~cm$^2$s$^{-1}$ at $\sim$1~GeV , a diffusion index of $\delta$~=~0.33, with fairly negligible convection and reacceleration~\citep{Trotta:2010mx}), the diffusion of cosmic rays in the TeV halo is less efficient by nearly four orders of magnitude. We note that this does not require the propagation of particles to be diffusive, and in fact the spectrum of the Geminga TeV halo is best-fit if the particle propagation is convective (or is diffusive with a diffusion index $\delta$~=~0)~\cite{2017arXiv170208436H}. 

Second, the slow propagation of particles, compared to the $\sim$60~yr light-crossing time of the TeV halo, indicates that the TeV halo is not beamed, but produces $\gamma$-ray emission isotropically. The isotropy of TeV halos implies that these systems could be detected even if the associated pulsar produces emission that is not beamed towards Earth, making the detection of TeV halos important for our understanding of the mis-aligned pulsar population.

 In order to determine the expected luminosity, spatial extent, and spectrum of these halos, we will assume throughout the remainder of this paper that the TeV halo luminosity of a pulsar is proportional to its spin-down luminosity, normalizing the ratio to Geminga:
 
 \begin{equation}
 \label{eq:fluxestimate}
 \phi_{{\rm TeV~halo}} = \left( \frac{\dot{E}_{{\rm psr}}}{\dot{E}_{{\rm Geminga}}}\right) \left(\frac{d^2_{{\rm Geminga}}}{d^2_{{\rm psr}}}\right)\phi_{{\rm Geminga}}
 \end{equation}
 
 \noindent We set $\phi_{{\rm Geminga}}$~=~4.9$\times$10$^{-14}$~TeV$^{-1}$cm$^{-2}$s$^{-1}$ at an energy of 7~TeV based on the best-fit HAWC value for the extended Geminga source~\citep{Abeysekara:2017hyn}. The value $\dot{E}$ is the spin-down luminosity of the pulsar, as calculated from the pulsar period and period derivative. The spin-down luminosity of Geminga is calculated in the ATNF catalog to be 3.2$\times$10$^{34}$~erg~s$^{-1}$.  The expected spatial extension of each source can be calculated as:
 
 \begin{equation}
 \label{eq:sizeestimate}
 \theta_{{\rm TeV~halo}} = \left(\frac{d_{{\rm Geminga}}}{d_{{\rm psr}}}\right) \theta_{{\rm Geminga}}
 \end{equation}
 
\noindent where we set $\theta_{{\rm Geminga}}$~=~2.0$^\circ$, again using the results provided by HAWC~\citep{Abeysekara:2017hyn}. We will discuss the nature of these HAWC observations in Section~\ref{sec:hawcsensitivity}.

\section{ACT Observations of TeV Halos}
\label{sec:acthalos}

Due to their large effective area and impressive angular resolution, ACTs have discovered most of the TeV halos observed to date. The most recent results from H.E.S.S. note that 19 TeV sources have been firmly associated with TeV halos\footnote{We note that in all cases, previous authors refer to these morphological features as TeV PWN. We maintain the phrasing of TeV halos throughout this paper for consistency.}, with an additional 20 potential associations~\cite{Abdalla:2017vci}. These observations indicate several important trends between observed TeV halos and the associated pulsar. First, a clear correlation is found between the pulsar spin-down power and the TeV halo luminosity, albeit with a dispersion that is $\sim$0.83~dex. Second, a correlation is found between the pulsar age and the radius of the TeV halo~\cite{Abdalla:2017vci}. This correlation is best presented in terms of the TeV halo radius and the spin-down power (which is strongly correlated with the pulsar age), and is found to scale as R$_{{\rm Halo}}$~$\propto$~$\dot{E}^{-0.65 \pm 0.20}$.

Most importantly, these results indicate that the population of H.E.S.S. detected TeV halos is strongly biased towards distant sources with high spin-down powers. Of the 35 pulsars associated with TeV halos\footnote{Four pulsars in the H.E.S.S. catalog are associated with two possible TeV sources.}, only five have spin-down powers below 10$^{36}$~erg~s$^{-1}$, and only two have ages above 100~kyr. This biased population of TeV halos is expected based on two factors affecting the H.E.S.S. sensitivity. First, since the H.E.S.S. sample is flux-limited, we expect that at any radial distance there is a minimum spin-down power $\dot{E}_{min}$~$\propto$~r$^2$, below which a TeV halo is too dim to be seen. Thus flux-threshold is the standard expectation for any class of sources. In the case of H.E.S.S. observations, there is a second sensitivity limit due to the fact that H.E.S.S. observations are insensitive to any TeV halos that are spatially extended by more than $\sim$0.6$^\circ$~\cite{Abdalla:2017vci}. Since the radius of a TeV halo varies inversely with the spin-down power, as shown above, this can be translated to a second sensitivity limit which is given by $\dot{E}_{min}$~$\propto$~r$^{-1.53}$. Note that these two limits have opposite slopes, and thus combine to strongly limit the parameter space of H.E.S.S. observations. ACTs are only sensitive to systems with extremely high spin-down powers found at moderate distances from Earth. In particular, given the best-fit models for these correlations calculated by~\cite{Abdalla:2017vci}, H.E.S.S would be incapable of detecting any TeV halo with a luminosity below 10$^{36}$~erg~s$^{-1}$ located within $\sim$2.4~kpc of Earth. 

\section{HAWC Observations of TeV Halos}
\label{sec:hawcsensitivity}

\begin{table*}[t]

\begin{tabular}{| c | c | c | c | c | c | c | c | c | c | c | c |}
\hline\hline
2HWC & ATNF & Distance & Angular & Projected & Expected & Actual & Flux & Expected & Actual & Age & Chance\\
Name & Name & (kpc) & Separation & Separation & Flux ($\times10^{-15}$) & Flux ($\times10^{-15}$) & Ratio & Extension & Extension & (kyr) & Overlap\\ \hline
J0700+143 & B0656+14 & 0.29 & 0.18$^\circ$ & 0.91~pc & 43.0 & 23.0 & 1.87 & 2.0$^\circ$ & 1.73$^\circ$ & 111 & 0.0 \\ \hline
J0631+169 & J0633+1746 & 0.25 & 0.89$^\circ$ & 3.88~pc  & 48.7 & 48.7 & 1.0 & 2.0$^\circ$ & 2.0$^\circ$ & 342 & 0.0 \\ \hline
J1912+099 & J1913+1011 & 4.61 & 0.34$^\circ$ & 27.36~pc & 13.0 & 36.6 & 0.36 & 0.11$^\circ$ & 0.7$^\circ$ & 169 & 0.30\\ \hline\hline
J2031+415 & J2032+4127 & 1.70 & 0.11$^\circ$ & 3.26~pc & 5.59 & 61.6 & 0.091 & 0.29$^\circ$ &  0.7$^\circ$ & 181 & 0.002\\ \hline
J1831-098 & J1831-0952 & 3.68 & 0.04$^\circ$ & 2.57~pc & 7.70 &  95.8 & 0.080 & 0.14$^\circ$ & 0.9$^\circ$ &128 & 0.006 \\\hline
\hline
\end{tabular}
%\footnote{Fluxes are recorded using the 2HWC convention, which lists the differential flux at 7~TeV in units TeV$^{-1}$~s$^{-1}$~cm$^{-2}$.}
%\footnote{2HWC J0631+169 is Geminga, thus the Ratio is unity by construction. We additionally note that while Geminga has been detected in radio observations, its radio intensity is extremely weak and its detection has relied mostly on its bright X-ray and $\gamma$-ray emission~\citep{1997Natur.389..697M}.}
\caption{HAWC sources listed in the 2HWC that are associated, or potentially associated, with an ATNF pulsar of age greater than 100~kyr. These systems have the highest probability of being TeV halos. This source list is meant to be maximally inclusive, including both potential chance associations, and sources for which the majority of the TeV emission may come from an associated supernova remnant. For each source, we list the distance as estimated by the ATNF catalog, along with the angular separation and projected separation between the 2HWC source and the ATNF pulsar. In addition, we provide the flux and spatial extension expected if each pulsar were represented as a Geminga-like pulsar (same efficiency in converting spin-down power into e$^+$e$^-$ production, see Equations~\ref{eq:fluxestimate}~and~\ref{eq:sizeestimate}). These predictions are compared to the actual flux and extension reported in 2HWC.  The fluxes are recorded following 2HWC convention, which lists the differential flux at 7~TeV in units of TeV$^{-1}$~s$^{-1}$~cm$^{-2}$. 
The ratio is defined as the expected flux divided by the actual flux. The quoted age is the characteristic age, $P$/2$\dot{P}$, and is an approximation of the true pulsar age. Finally, we list the probability that a random ATNF pulsar will fall in the region bounded by the angular offset between the 2HWC source and the ATNF pulsar, as described in the text. The two systems listed under the double horizontal line are tenuous associations, as the projected TeV halo flux is more than an order of magnitude smaller than the observed flux from the system. 2HWC J0631+169 is Geminga, thus the ratio is unity by construction.}
\label{tab:HAWCsources}
\end{table*}

\begin{table*}[t]
	
\begin{tabular}{| c | c | c | c | c | c | c | c | c | c | c | c |}
\hline\hline
2HWC & ATNF & Distance & Angular & Projected & Expected & Actual & Flux & Expected & Actual & Age & Chance\\
Name & Name & (kpc) & Separation & Separation & Flux  ($\times10^{-15}$) & Flux  ($\times10^{-15}$) & Ratio & Extension & Extension & (kyr) & Overlap \\ \hline
J1930+188 & J1930+1852 & 7.0 & 0.03$^\circ$ & 3.67~pc & 23.2 & 9.8 & 2.37 & 0.07$^\circ$ & 0.0$^\circ$ & 2.89 & 0.002\\ \hline
J1814-173 & J1813-1749 & 4.7 & 0.54$^\circ$ & 44.30~pc & 243 & 152 & 1.60 & 0.11$^\circ$ & 1.0$^\circ$  & 5.6 & 0.61\\ \hline
J2019+367 & J2021+3651 & 1.8 & 0.27$^\circ$ & 8.48~pc & 99.8 & 58.2 & 1.71 & 0.28$^\circ$ & 0.7$^\circ$ & 17.2 & 0.04\\ \hline
 J1928+177 & J1928+1746 & 4.34 & 0.03$^\circ$ & 2.27~pc & 8.08 & 10.0 & 0.81 & 0.11$^\circ$ & 0.0$^\circ$ & 82.6 & 0.002\\ \hline 
J1908+063 & J1907+0602 & 2.58 & 0.36$^\circ$ & 16.21~pc & 40.0 & 85.0 & 0.47 & 0.2$^\circ$ & 0.8$^\circ$ & 19.5 & 0.26\\ \hline
 J2020+403 & J2021+4026 & 2.15 & 0.18$^\circ$ & 6.75~pc & 2.48 & 18.5 & 0.134 & 0.23$^\circ$ & 0.0$^\circ$ & 77 & 0.01 \\ \hline
J1857+027 & J1856+0245 & 6.32 & 0.12$^\circ$ & 13.24~pc & 11.0 & 97.0 & 0.11 & 0.08$^\circ$ & 0.9$^\circ$ & 20.6 & 0.06 \\ \hline\hline
J1825-134 & J1826-1334 & 3.61 & 0.20$^\circ$ & 12.66~pc & 20.5 & 249 & 0.082 & 0.14$^\circ$ & 0.9$^\circ$ & 21.4 & 0.14\\ \hline
J1837-065 & J1838-0655 & 6.60 & 0.38$^\circ$ & 43.77~pc & 12.0 & 341 & 0.035 & 0.08$^\circ$ & 2.0$^\circ$  & 22.7 & 0.48 \\ \hline
J1837-065 & J1837-0604 & 4.78 & 0.50$^\circ$ & 41.71~pc & 8.3 & 341 & 0.024 & 0.10$^\circ$ & 2.0$^\circ$  & 33.8 & 0.68 \\\hline
 J2006+341 & J2004+3429 & 10.8 & 0.42$^\circ$ & 80.07~pc & 0.48 & 24.5 & 0.019 & 0.04$^\circ$ & 0.9$^\circ$ & 18.5 & 0.08\\ \hline
\end{tabular}
%\footnote{Fluxes are recorded using the 2HWC convention, which lists the differential flux at 7~TeV in units TeV$^{-1}$~s$^{-1}$~cm$^{-2}$.}
%\footnote{2HWC J0631+169 is Geminga, thus the Ratio is unity by construction. We additionally note that while Geminga has been detected in radio observations, its radio intensity is extremely weak and its detection has relied mostly on its bright X-ray and $\gamma$-ray emission~\citep{1997Natur.389..697M}.}
\caption{Same as Table~\ref{tab:HAWCsources} for 2HWC sources correlated with pulsars that have characteristic ages below 100~kyr. These systems (compared to those in Table~\ref{tab:HAWCsources}) are more likely to be contaminated by considerable emission from an affiliated supernova remnant. Moreover, their age is similar to the cooling time of TeV e$^+$e$^-$ making their luminosity uncertain. We note that the characteristic age, $P$/2$\dot{P}$, is approximate, and typically overestimates the age of the youngest pulsars. 2HWC J1837-065 is potentially associated with two ATNF pulsars.}
\label{tab:HAWCsourcesyoung}
\end{table*}

In this section, we argue that HAWC observations probe an important new parameter space not observable by ACTs, namely the population of nearby middle-aged pulsars with significantly extended TeV halos. Unlike ACTs, HAWC directly detects the particles within the air showers generated by $\gamma$-rays, allowing it to simultaneously observe a large field-of-view ($>$1.5~sr). At present, HAWC is the best tool to detect very high energy \mbox{$\gamma$-rays} from spatially extended sources. 

At present, HAWC has accumulated $\sim$17 months of observations, achieving a remarkable sensitivity of 5-10\% Crab~\cite[hereafter 2HWC]{Abeysekara:2017hyn}. The exact sensitivity varies based on the declination of the observed source, with a maximum sensitivity at $b$~=~20$^\circ$ that decreases by a factor of $\lesssim$~2 for sources with declinations differing by up to $\sim$30$^\circ$. The HAWC sensitivity is computed as a differential flux at a standard energy of $\sim$7~TeV, and in these units it varies significantly as a function of the assumed spectral index. At $b$~=~20$^\circ$, the flux sensitivity varies from (3 -- 6)$\times$10$^{-15}$~TeV$^{-1}$cm$^{-2}$s$^{-1}$ at 7~TeV, for assumed spectral indices of -2.0 and -2.5, respectively. 

We compare these sensitivities to the detected fluxes of the spatially extended TeV halos surrounding both Geminga and Monogem, which are (4.87~$\pm$~0.69)$\times$10$^{-14}$~TeV$^{-1}$cm$^{-2}$s$^{-1}$ and (2.30~$\pm$~0.73)$\times$10$^{-14}$~TeV$^{-1}$cm$^{-2}$s$^{-1}$, with spectral indexes of \mbox{-2.23~$\pm$~0.08} and \mbox{-2.03~$\pm$~0.14}, respectively~\cite{Abeysekara:2017hyn}. We assume that the HAWC sensitivity for these sources is the average of the quoted sensitivities for spectral indices of -2.5 and -2.0, and calculate that Geminga (Monogem) would be observed out to a distance of $\sim$950~pc ($\sim$650~pc) if it were found at the optimal declination of $b$~=~20$^\circ$. Geminga- or Monogem-like pulsars observed throughout the range -10$^\circ$~$<$~b~$<$~50$^\circ$ would be observed to distances exceeding 660~pc (450~pc). These values significantly exceed the observed distances to each pulsar of 250$^{+230}_{-80}$~pc (280$^{+30}_{-30}$~pc)~\citep{2012ApJ...755...39V}\footnote{For the remainder of this paper, we will utilize the ATNF distance measurements of 250~pc (290~pc) for Geminga (Monogem), in order to facilitate comparisons with the remainder of the ATNF catalog.}. If the HAWC sensitivity scales with the square root of time, 10 years of data will be sensitive to emission from similar pulsars up to distances of $\sim$1.5~kpc.

The above estimate does not account for the degradation in HAWC sensitivity for extended emission sources. The HAWC sensitivity to extended TeV halos will depend on the size of the emission source as well as the modeling of both diffuse $\gamma$-ray and cosmic-ray backgrounds. However, early results in the 2HWC are promising. Already the source 2HWC~J1040+308 was observed with a statistically significant spatial extent of 0.5$^\circ$, and an observed flux of \mbox{(6.6~$\pm$~3.5)~$\times$~10$^{-15}$~TeV$^{-1}$~cm$^{-2}$~s$^{-1}$}. This flux lies only a factor of $\sim$2 above the current point-source sensitivity of HAWC. Moreover, distant TeV halos will have smaller angular extensions, producing sensitivities closer to the nominal values for point-source emission.

Intriguingly, Monogem and Geminga are not unique in the 2HWC catalog. Of the 39 2HWC sources, seven are listed as possibly associated with a PWN~\cite{Abeysekara:2017hyn}. Additionally, 18 sources are found to exhibit detectable spatial extension. These 2HWC sources appear similar, but not identical, to the $\sim$20\% of TeV catalog sources observed primarily by ACTs that are currently unidentified\footnote{http://tevcat.uchicago.edu/}. Some combination of these sources may comprise a significant population of extended TeV halos that produce a significant fraction of the total TeV $\gamma$-ray sky.

\subsection{Current Tentative Detections} 

\begin{table*}[t]
\begin{tabular}{| c | c | c | c | c | c | c | c | c |}
\hline\hline
ATNF Name & Dec. ($^\circ$) & Distance (kpc) & Age (kyr) & Spindown Lum. (erg s$^{-1}$) & Spindown Flux (erg s$^{-1}$~kpc$^{-2}$) & 2HWC\\\hline
J0633+1746 & 17.77 &  0.25 &  342 & 3.2e34 & 4.1e34 & 2HWC J0631+169  \\ \hline
B0656+14  &  14.23 &  0.29 &  111 & 3.8e34 & 3.6e34 & 2HWC J0700+143  \\  \hline
B1951+32  &  32.87 &  3.00 &  107 & 3.7e36 & 3.3e34 & --- \\ \hline
J1740+1000 & 10.00 &  1.23 &  114 & 2.3e35 & 1.2e34 & --- \\ \hline
J1913+1011 & 10.18 &  4.61 &  169 & 2.9e36 & 1.1e34 & 2HWC J1912+099  \\ \hline
J1831-0952 & -9.86 &  3.68 &  128 & 1.1e36 & 6.4e33 & 2HWC J1831-098  \\ \hline
J2032+4127 & 41.45 &  1.70 &  181 & 1.7e35 & 4.7e33 & 2HWC J2031+415  \\ \hline
B1822-09  &  -9.58 &  0.30 &  232 & 4.6e33 & 4.1e33 & --- \\ \hline
B1830-08  &  -8.45 &  4.50 & 147 & 5.8e35 & 2.3e33 & --- \\ \hline
J1913+0904 & 9.07  & 3.00  & 147 & 1.6e35 & 1.4e33 & --- \\ \hline
B0540+23 &   23.48  & 1.56 &  253 & 4.1e34 & 1.4e33 & --- \\ \hline
\hline
\end{tabular}
\caption{The 11 ANTF catalog sources with ages between 100-400~kyr that are located in a declination range accessible to HAWC and have expected TeV halo fluxes that are at least 2\% as large as the measured Geminga flux (assuming an equivilent conversion efficiency of spin-down power to e$^+$e$^-$ pairs in all systems). The distance to each source is based on the calculated free-electron density~\citep{Cordes:2002wz}, and the spin-down luminosity is the value reported in the ATNF catalog. The spin-down flux is calculated from the spin-down luminosity and distance. We provide the 2HWC name for sources potentially associated with HAWC catalog sources. We note that Geminga and Monogem are expected to be the brightest TeV halos observable by HAWC, and three of the next five brightest systems have already been detected in current HAWC observations. The current spin-down flux sensitivity of HAWC should be $\sim$4$\times$10$^{33}$~erg~s$^{-1}$~kpc$^{-2}$, with significant uncertainties.}
\label{tab:atnfsources}
\end{table*}

In Tables~\ref{tab:HAWCsources}~and~\ref{tab:HAWCsourcesyoung} we list 15 2HWC sources\footnote{2HWC J1837-065 is potentially associated with two separate ATNF pulsars. Thus, there are 15 associated 2HWC sources, and 16 possible associated pulsars.} possibly associated with ATNF (Australia Telescope National Facility~\citep{2005AJ....129.1993M}\footnote{http://www.atnf.csiro.au/people/pulsar/psrcat/}) radio pulsars (including Geminga and Monogem). Seven of these systems have been labeled as possible associations in the 2HWC catalog, while nine others are listed based on their angular proximity to a known pulsar. We have aimed to be maximally inclusive in producing these lists. We have separated these sources into two tables based on the age of the associated pulsar, noting that younger pulsars are less likely to be in steady state, and more likely to be significantly contaminated by emission from the associated supernova remnant. 

From these tables we note three interesting results. First, a significant fraction of the 39 2HWC sources are located near an ATNF radio pulsar. As there are $\sim$2500 ATNF radio pulsars, and the majority of 2HWC sources lie in the Galactic plane, it is possible that some of these associations are due to chance overlaps. To calculate the number of expected chance coincidences for each 2HWC source, we count the average number of ATNF pulsars with characteristic ages below 10$^6$~yr in a 20$^\circ\times$2$^\circ$ strip in longitude and latitude centered on each pulsar association. Assuming that these sources were repositioned randomly in the region, we calculate the probability that a source would lie within the circular region determined by the angular separation between the 2HWC source and the ATNF pulsar. We find that of the 16 listed associations, only 2.67 are expected to be due to chance associations. Moreover, these chance associations are dominated by the double association of 2HWC J1837-065 along with the system 2HWC J1814-173. The number of chance overlaps among the remaining 13 systems is only 0.90. Eight systems have a smaller than 5\% probability of being explained as a chance overlap. This strongly indicates that a large fraction of all 2HWC sources are associated with pulsar activity. However, this does not preclude the probability that these systems are TeV bright due to a convolving factor, such as the supernova remnant that is associated with the pulsar. 

Second, 9 of these 14 systems\footnote{Ignoring Geminga, which is Geminga-like by definition.} have an observed flux that falls within an order of magnitude of the expected flux from a Geminga-like TeV halo. For sources that are nearly an order of magnitude brighter than the Geminga-like expectation, the TeV halo interpretation may be stretched. Notably, the intensity of the TeV halo surrounding Geminga requires that \mbox{7-29\%} of the spin-down power is converted into e$^+$e$^-$~\citep{2017arXiv170208436H}. Any TeV halo that exceeds the Geminga-like flux by an order of magnitude would require a pair-conversion efficiency that approaches or exceeds unity. However, large variations are expected in the modeled TeV flux due to uncertainties in the distance, e$^+$e$^-$ spectrum, and local interstellar radiation field of each pulsar. Additionally, the power that is injected into e$^+$e$^-$ should be compared to the average spin-down power over the $\sim$20~kyr cooling time of 10~TeV e$^+$e$^-$ in the TeV halo. Utilizing the current spin-down power of each pulsar potentially underestimates the energy available for e$^+$e$^-$ injection. Finally, we note that the 2HWC catalog is highly biased by the HAWC sensitivity cut, and the observed systems are likely to be those with the largest upward fluctuation in their flux compared to the average TeV source.

Third, the majority of 2HWC sources shown in Tables~\ref{tab:HAWCsources}~and~\ref{tab:HAWCsourcesyoung} are coincident with relatively young pulsars (11 have characteristic ages below 100~kyr). This echos previous observations by H.E.S.S.~\cite{Abdalla:2017vci}, which finds a large population of TeV halos coincident with young pulsars. This is also expected theoretically, as young systems have extremely high spin-down powers.  Within the context of our Geminga-like model, these systems are expected to provide a tantalizing population of highly luminous TeV halos. However, in what follows, we will conservatively ignore the contribution from systems with characteristic ages below 100~kyr for three reasons. First, their TeV emission is more likely to be contaminated by bright emission from their corresponding supernova remnant, making the fractional contribution of the TeV halo to the total $\gamma$-ray emission difficult to determine. Second, they may not be in steady state, as their age may be smaller than the e$^+$e$^-$ cooling time. Third, they are less likely to exhibit significant spatial extension, as the size of the TeV halo (like the X-ray PWN) is expected to expand over time (see Equation~\ref{eq:pwnradius}). However, in Section~\ref{sec:youngpwn} we will integrate these sources into our model, and consider several observational tests that can be performed using the joint HAWC and H.E.S.S. catalogs.

We stress that several associations in this list are tenuous, and we intend Tables~\ref{tab:HAWCsources}~and~\ref{tab:HAWCsourcesyoung} to err on the side of inclusivity. In particular, many ATNF pulsars appear coincident with bright (and likely associated) supernova remnants, which may contribute the majority of the TeV emission. The source 2HWC J1837-065 is potentially associated with two different ATNF pulsars. We note that several of the 2HWC sources (most notably 2HWC J1837-065 and 2HWC J1857+027) have observed spatial extensions which exceed that expected from a Geminga-like system by more than an order of magnitude. These may be difficult to accommodate within our model of TeV halos, though the expected spatial extension depends sensitively on assumed diffusion coefficients within the TeV halo. Finally, as we will discuss in Section~\ref{sec:youngpwn}, if all of the sources listed in Tables~\ref{tab:HAWCsources}~and~\ref{tab:HAWCsourcesyoung} are TeV halos associated with known pulsars, the number of TeV halos produced by currently unknown pulsars would exceed the 39 observed 2HWC sources.

\subsection{Predicted Detections}
Using Geminga as a standard candle for TeV halos, we can predict which ATNF radio sources are most likely to be associated with bright TeV halos. In Table~\ref{tab:atnfsources}, we provide a list of the 11 ATNF radio pulsars that fit the following criteria: (1) an age between 100-400~kyr, (2) a declination in the HAWC field-of-view (between -10$^\circ$ and 50$^\circ$), and (3) an expected flux exceeding 1.0$\times$10$^{33}$~$\eta$~erg~s$^{-1}$~kpc$^{-2}$, where $\eta$ is the (assumed universal) efficiency in converting spin-down power into TeV halo emission. These systems are expected to have fluxes exceeding $\sim$1$\times$10$^{-15}$~TeV$^{-1}$cm$^{-2}$s$^{-1}$ at 7~TeV, and to eventually be detectable by HAWC. Implementing the 100-400~kyr age cut significantly decreases the population of systems compared to those shown in Tables~\ref{tab:HAWCsources}~and~\ref{tab:HAWCsourcesyoung}, and limits the potential overlap of our model with bright supernova remnants.

Intriguingly, five of the seven ATNF pulsars with the brightest expected TeV halos are associated with a 2HWC source. Moreover, all five of the middle-aged pulsars associated with 2HWC sources in Table~\ref{tab:HAWCsources} were expected to be among the brightest TeV halos. As there are 55 ATNF sources corresponding to middle-aged pulsars in the HAWC field, this overlap strongly suggests a close correlation between the pulsar spin-down luminosity and the luminosity of the TeV halo.

We note some tension with the pulsars B1951+32 and J1740+1000. Using our Geminga-like model, these systems are expected to have fluxes of 3.9$\times$10$^{-14}$~TeV$^{-1}$cm$^{-2}$s$^{-1}$ and 1.4$\times$10$^{-14}$~TeV$^{-1}$cm$^{-2}$s$^{-1}$ respectively, and should be detectable in the 2HWC. The current HAWC upper limit for the flux from these point sources is not known. However, the lack of detected TeV halos from these systems indicates that there is either dispersion in the value of $\eta$, or alternatively that the diffusion environment of TeV halos differs significantly between systems. This may decrease the number of observable TeV halos by a factor of $\sim$2. We will discuss this in more detail in Section~\ref{sec:discussion}.

\begin{figure*}[tbp]
\centering
\includegraphics[width=.95\textwidth]{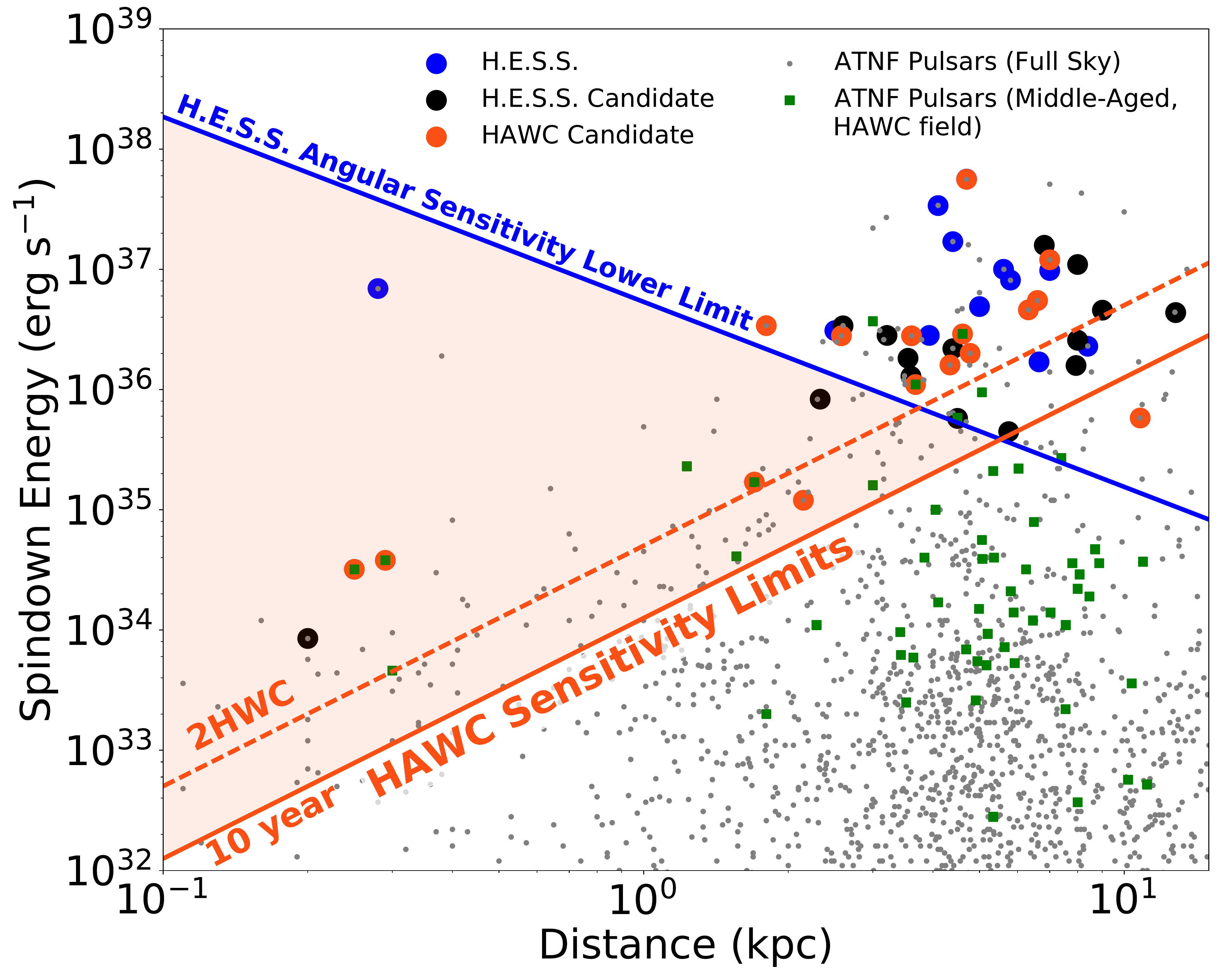}
\caption{The new discovery space provided by HAWC observations of TeV halos in the Milky Way (shaded orange region). Orange points represent 2HWC sources associated with ATNF pulsars, as listed in Tables~\ref{tab:HAWCsources}~and~\ref{tab:HAWCsourcesyoung}. Blue (black) datapoints represent H.E.S.S. TeV halos associated (potentially associated) with ATNF pulsars, as provided by~\cite{Abdalla:2017vci}. Gray datapoints represent ATNF radio pulsars with known distances and spin-down energies~\cite{2006MNRAS.372..777L}. The gray circles include all ATNF pulsars, regardless of whether they lie within the field-of-view of HAWC. Thus, we stress that isolated gray points above the HAWC sensitivity threshold do not indicate failed detections. The green squares include only middle-aged ATNF pulsars that lie within the HAWC field-of-view. The orange dashed (solid) line represents the sensitivity of HAWC in the 2HWC catalog (after 10 years of observation), assuming all pulsars produce TeV halos with luminosities calculated using our Geminga-like model. The blue H.E.S.S. angular sensitivity lower limit excludes regions of parameter space where the TeV halo would be expected to be extended by more than 0.6$^\circ$~\cite{Abdalla:2017vci}. H.E.S.S. observations also include a flux sensitivity limit (not shown), which falls within a factor of $\sim$2 of the HAWC 10~year sensitivity limit, depending on the H.E.S.S. observation time. The large number of gray datapoints in the HAWC sensitivity region (orange shaded) demonstrate the potential for HAWC to observe a large number of new TeV halos.}
\label{fig:hawcdetectionspace}
\end{figure*}

In Figure~\ref{fig:hawcdetectionspace} we combine our results from the previous two sections, and demonstrate that HAWC observations open a vast new parameter space for TeV halo detections --- middle-aged TeV halos in close proximity to Earth. To illustrate this parameter space, we show all TeV halo associations and ATNF pulsars, regardless of their characteristic age or location relative to the HAWC field of view. We first note that bright pulsars at large distances are efficiently detected by both HAWC and ACTs. In fact, a significant fraction of all known ATNF pulsars with high spin-down periods have been detected at TeV energies. This indicates that TeV halos are a generic feature of pulsars. 

However, the small field-of-view of ACTs inhibits the detection of TeV halos with radial extents exceeding $\sim$0.6$^\circ$. We utilize the correlation between the radial extent of a TeV halo and its spin-down power~\cite{Abdalla:2017vci}, and find that this prevents ACTs from observing TeV halos produced by pulsars with spin-down luminosities below 10$^{36}$~erg~s$^{-1}$ at distances below $\sim$2.5~kpc. The wide field-of-view of HAWC allows us to detect these systems. ATNF observations indicate that this region of parameter space (depicted as an orange shaded region) includes approximately twice as many pulsars as the region accessible to ACTs. We note that pulsars in the bottom right corner are inaccessible to both ACTs and HAWC due to their very small TeV fluxes.

\section{Incompleteness in the Nearby Pulsar Population}
\label{sec:sourcenumbers}

In Section~\ref{sec:hawcsensitivity}, we demonstrated that HAWC observations have the unique ability to observe TeV halos from nearby pulsars with low spin-down power. However, thus far we have only considered systems coincident with known ATNF radio pulsars. We now discuss the capability of HAWC observations to detect a large population of ``invisible" pulsars without any ATNF association. 

Stars with initial masses between $\sim$8 - 25~M$_\odot$ are expected to end their lives as neutron stars~\citep{1984ApJ...277..791N, 1999ApJ...522..413F}. A fraction (potentially unity) of these neutron stars will move through a pulsar stage that is expected to last $\mathcal{O}$(100~Myr). During this period, strong magnetic fields on the pulsar surface~\citep{1971ApJ...164..529S} and termination shocks in the pulsar wind nebula accelerate e$^+$e$^-$ to extremely high energies~\citep{1998PhRvL..80.3911B}. As the pulsar slows down, these fields decay and particle acceleration ceases. 

To date, over 2500 pulsars have been detected using their beamed radio emission (see e.g.~\citep{2005AJ....129.1993M}). This constitutes the vast majority of known pulsars, most of which have not been detected at other wavelengths. Thus, the known pulsar population is highly biased towards systems with radio beams oriented towards Earth. The fraction of pulsars with favorable radio beam orientations is modeled by~\citep{1998MNRAS.298..625T}:

\begin{equation}
\label{eq:beaming}
f = \left[1.1~\left({\rm log}_{10}\left(\frac{\tau}{100~{\rm Myr}}\right)\right)^2 + 15\right]\%
\end{equation}

\noindent For middle-aged pulsars (100-400~kyr) this corresponds to beaming fractions between 21--25\%. We note that these results are based on a braking index of 3, and may change by $\sim$50\% (i.e. 10\% - 30\%) for reasonable modifications of the time evolution of the braking index. To approximate the number of mis-aligned pulsars with TeV halos detectable by HAWC, we sum the inverse of the beaming fraction of each ATNF pulsar listed in Table~\ref{tab:atnfsources}. The 11 ATNF radio sources listed in Table~\ref{tab:atnfsources} are likely to be the detectable subset of a population that includes $\sim$48 pulsars. Using a binomial distribution with the number of detected pulsars set to 11, we estimate that 37$^{+17}_{-13}$ additional ``invisible" pulsars exist. These pulsars have intrinsic characteristics similar to those listed in Table~\ref{tab:atnfsources}, but have radio beams that are not oriented towards Earth. We note that this is a statistical uncertainty, and that the actual uncertainty in the implied population is larger, given the uncertainties associated with the beaming fraction of Equation~\ref{eq:beaming}. 

Of the 11 sources listed in Table~\ref{tab:atnfsources}, five are consistent with TeV sources in the 2HWC catalog. Using binomial statistics once again to estimate the underlying population of TeV halos, we find that the current observation of HAWC sources surrounding Geminga, Monogem, 2HWC J1912+099, 2HWC J1831-098, and 2HWC J2031+415 would indicate that an additional population of 16$^{+12}_{-9}$ TeV halos invisible to radio observations should already be observed in the 2HWC catalog. This is potentially problematic, given that only 27 2HWC sources are currently not associated with low-energy emission, and Tables~\ref{tab:HAWCsources}~and~\ref{tab:HAWCsourcesyoung} indicate that many TeV halo candidates correspond to young pulsars that are not in our sample. This result implies that the current observation of five TeV halos in Table~\ref{tab:atnfsources} is either due to a slightly fortunate arrangement of beaming angles among the brightest TeV halos, or alternatively that one or two of the 2HWC sources listed in Table~\ref{tab:atnfsources} is not, in fact, a TeV halo.

Additionally, the above calculation is conservative. While the ATNF catalog lists 53 middle-aged pulsars within the HAWC field-of-view, an additional 155 non-millisecond ATNF pulsars without defined ages are also found in this region. Of these 155 systems, 36 are located within 2~kpc of the Earth, indicating that they may be among the brightest pulsars, depending on their spin-down luminosity. Of these 36 systems, 20 have rotation periods below 1~s and 10 have rotation periods below 0.5~s, which are very roughly compatible with a middle-aged pulsar population. However, it is difficult to estimate how many of these systems have spin-down luminosities indicative of a bright pulsar population. 

As an alternative estimate for the number of middle-aged pulsars near Earth, we employ a pulsar distribution following the Lorimer parametrization \cite{Lorimer:2003qc}. Specifically, we calculate the pulsar column density as \mbox{$\rho_r \propto r^{n}\exp \{-r/\sigma \}$} where $r$ is the galactocentric distance and $n$ and $\sigma$ are fit parameters. We transform this function to a pulsar surface density around the solar position. We then fit $n$ and $\sigma$ to the observed number of pulsars with an age of up to 10$^7$~yr within 5.5 kpc from the solar position. We account for the fact that more distant pulsars are only detectable if they are particularly bright by including a relative normalization factor of \mbox{$1/(1+r^\alpha)$}, where $\alpha$ is a fit parameter. We normalize the total number of pulsars in the Milky Way with an age of up to 10$^7$~yr to be 2$\times$10$^5$, based on a birth rate of 2 pulsars per century (for details \cite{Karwal:2017}).  We obtain best-fit values of $n$=2.1 and $\sigma$=1.14 kpc. These models predict a beaming fraction of $\sim$20\% for middle-aged pulsars, in agreement with previous estimates~\citep{1998MNRAS.298..625T}. 

Since this model is insensitive to the beaming fraction of the pulsar population, we can directly calculate the number of expected pulsars as a function of the solar distance. Our model indicates an expected number of 13 (60) middle-aged pulsars within 1~kpc (2~kpc) of the Sun. We compare this population to the 9 (19) middle-aged ATNF pulsars (with favorable beaming angles) within 1~kpc (2~kpc) of the Sun. Given that the beaming fraction for middle-aged pulsars is expected to be $\sim$20-25\%, we find that the number of nearby pulsars appears to exceed the predicted value by nearly a factor of two. In part, this is likely due to the presence of the spiral arms, which produce significant over-densities within a kpc of the Sun and are not accounted for in the Lorimer parameterization. On the other hand, the number of observed pulsars between \mbox{1-2~kpc} from the Earth (47) is compatible with the ATNF population and a beaming fraction given by Equation~\ref{eq:beaming}. We note, however, to make this comparison work the population of 36 pulsars without known ages must contain few middle-aged systems.

\subsection{Multi-Wavelength Detections of Misaligned Pulsars}

While most pulsars have been detected via their beamed radio emission, it is possible that some nearby pulsars have been detected through either their $\gamma$-ray pulsations or their X-ray PWN. This would decrease the number of TeV halos detected by HAWC that are not associated with known emission sources. Unlike the case of radio observations, where distance measurements can be made based on pulse dispersion, the proximity of these sources will remain unknown. However, large $\gamma$-ray or PWN fluxes would be indicative of bright TeV halo emission as shown in Equation~\ref{eq:fluxestimate}. 

We first consider Fermi-LAT pulsars that were detected in blind $\gamma$-ray searches. The Fermi Second Pulsar Catalog (along with updates available online)~\citep{2013ApJS..208...17A} contains 107 young (non-millisecond) pulsars, including 54 selected by blind-search $\gamma$-ray observations. However, out of these 54 systems, 35 have characteristic ages that fall outside of the 100-400~kyr timescale employed throughout this paper, and 12 more have no age determination. Of the 19 systems which might be middle-aged, only five (J1844-0346 J0622+3749, J2017+3625, J1846+0919, and J2032+4127) fall within the latitude range observable by HAWC, three of which are known to be middle-aged (J0622+3749,  J1846+0919, and J2032+4127). The first two systems have spin-down luminosities below 3$\times$10$^{34}$~erg~s$^{-1}$, and would thus only be observable (with 10~yr observations) if they fell within $\sim$1.4~kpc of Earth. J2032+4127 has a spin-down luminosity of 3$\times$10$^{35}$~erg~s$^{-1}$, and is potentially observable out to $\sim$4.5~kpc from Earth. Still, these systems encompass only a small number of the missing TeV halo population. Thus we conclude that currently detected Fermi-LAT pulsars constitute only a small fraction of the expected population of TeV halos.

On the other hand, observations of Geminga and PSR J1954+2836 indicate the potential for $\gamma$-ray pulsars to be correlated with bright TeV halos. While Geminga has been detected as a radio pulsar, it is an extremely dim radio emitter. The detection of Geminga depends on its bright X-ray and $\gamma$-ray emission, and it is not clear that the pulsar would have been detected in blind radio searches~\citep{1997Natur.389..697M}. The $\gamma$-ray pulsar J1954+2836 has a characteristic age of only 69~kyr, and thus does not make the age cut for ATNF pulsars listed in Table~\ref{tab:atnfsources}. However, this pulsar is potentially correlated with the HAWC source 2HWC J1955+285. While the distance to PSR J1954+2836 is not known, it has a large spin-down luminosity of 1.0$\times$10$^{36}$~erg~s$^{-1}$, making it potentially observable by HAWC at distances up to $\sim$7~kpc from Earth.

Mis-aligned pulsars could also be detected via an associated X-ray PWN. Since PWN emit isotropically, the population of detected X-ray PWN should significantly exceed the population of detected pulsars. At present, however, the catalogs of PWN without associated pulsars are sparse. An analysis by~\citep{2006csxs.book..279K} found 24 such sources. Subsequent  observations found  pulsars coincident with 10 of these PWN~\citep{2005AJ....129.1993M}. Of the remaining 14 sources, only 6 fall within the HAWC field-of-view. Moreover, the spin-down period of the pulsars found in follow-up observations indicate that these PWN tend to be much younger than 100~kyr.

A more recent effort directly considered the overlap between PWN and unidentified TeV sources~\citep{2013arXiv1305.2552K}, identifying a population of 15 additional PWN without known pulsars. Of these systems, five lie within the HAWC field-of-view, and only two lie within 5~kpc of Earth. However, one of these sources DA~495 (G65.73+1.18) does overlap with the HAWC source 2HWC J1953+294, indicating that this X-ray PWN likely has a TeV halo counterpart. Additionally, the second source IC~443 (G189.23+2.90) has potentially been detected in follow-up observations by VERITAS (VERITAS J0616.9+2230). This association is questionable, however, as IC 443 is known to be coincident with an extremely bright supernova remnant that likely produces the majority of its TeV emission~\citep{2010ApJ...712..459A}. These observations do indicate a strong association between X-ray PWN and TeV halos, and suggest that full sky observations of PWN candidates could potentially detect X-ray emission from many of the TeV halos observable by HAWC. 

By combining results from the ATNF pulsar catalog with theoretical models of the radio pulsar beam size, we estimate that 37$^{+17}_{-13}$ TeV halos may exist with fluxes detectable HAWC after 10 years of observation. Given that, at most, seven of these sources (and likely many fewer) have been detected by $\gamma$-ray or X-ray PWN observations, we are forced to conclude that a significant population of TeV halos may be initially detected by HAWC as unassociated TeV sources. Follow up observations of these sources will be necessary to determine their TeV halo origin.

\section{Follow-Up Observations of TeV Halos}
\label{sec:pwn}

In the previous section, we established that HAWC can observe a sizable population of new TeV halos. Here, we argue that multiwavelength observations can determine the nature of these sources. This argument is primarily motivated by two facts. First, our models predict that a large fraction of all 2HWC sources are produced by TeV halos. This implies that the rate of false positives is low and that follow-up observations are likely to be fruitful. Second, previous multiwavelength searches have been successful in the opposite direction. This signals that there is a close correlation between observations of pulsars, PWN and TeV halos. In particular, studies by~\citep{Aharonian:2005pv, 2013arXiv1305.2552K, Abdalla:2017vci} showed that a significant fraction of the brightest X-ray PWN have associated TeV emission that can be observed with targeted ACT searches.

\subsection{Diffuse X-ray Observations}

X-ray observations provide the clearest path forward. There are two morphological regions to consider. The first is diffuse X-ray emission with the same spatial extent as the TeV halo, a signature we will refer to as the ``X-ray halo". Since TeV halos are produced through the inverse-Compton scattering of ambient radiation, this emission should be coincident with synchrotron emission in the X-ray band. Thus, all TeV halos should produce morphologically similar X-ray halos. The ratio of the X-ray and TeV halo intensities depend only on the ratio of the magnetic field energy density to the ISRF energy density. Assuming a standard 5$\mu$G magnetic field and a 1.0~eV~cm$^{-3}$ ISRF, approximately half of the total e$^+$e$^-$ energy is lost in the form of synchrotron radiation. The spectrum of this synchrotron emission peaks near the critical energy:

\begin{equation}
E_{{\rm sync, critical}} =  22~{\rm eV} \left(\frac{B}{5~\mu G} \right) \left(\frac{E_e}{10~{\rm TeV}} \right)^2
\end{equation} 

\noindent Thus, the Chandra energy band ($>$0.2~keV) receives contributions primarily from electrons with energies exceeding $\sim$30~TeV. Utilizing a spectral fit for the Geminga pulsar described in~\citep{2017arXiv170208436H}, we calculate the synchrotron emission spectrum from an e$^+$e$^-$ injection spectrum of \mbox{$E^{-1.5}$~exp(-$E$/35~TeV)}. We assume energy-loss rates are dominated by synchrotron and inverse-Compton scattering. In this scenario, we find that only $\sim$3\% of the total e$^+$e$^-$ injection power above 1~TeV is converted to synchrotron radiation in the Chandra band. Normalizing this to the observed TeV halo flux, this equates to a count rate of $\sim$2~$\times$10$^{-6}$~ph~cm$^{-2}$~s$^{-1}$~deg$^{-2}$ above 200~eV, which is several orders of magnitude below current Chandra or XMM-Newton sensitivities. Thus, we conclude that it is unlikely that current X-ray observations could detect X-ray emission coincident with the TeV halo in a Geminga-like system.

However, X-ray halos could be detected in more luminous pulsars that are more distant than Geminga. These systems will have a significantly higher X-ray surface brightness due to their smaller angular size. Intriguingly, such a system may have already been detected coincident with the supernova remnant \mbox{G327.1-1.1~\citep{Temim:2015ula}}. This source, at an estimated distance of 9~kpc, boasts a ``Cometary PWN" of size $\sim$970 arcsec$^2$, surrounded by a ``Diffuse PWN" of size $\sim$20,000~arcsec$^2$. The size of the Diffuse PWN is approximately equivalent to the size of a Geminga-like TeV halo projected to a distance 9~kpc from Earth. Additionally, this source is known to produce bright TeV emission detectable by H.E.S.S.~\citep{2011ICRC....7..185A}, and this TeV emission is found to be spatially extended and coincident with the PWN. This source could constitute the first joint detection of a TeV and X-ray halo. Unfortunately, this source does not fall within the HAWC field-of-view. 

Even if X-ray observations are not able to directly observe the synchrotron emission from X-Ray halos, X-ray observations will be more successful in detecting the compact PWN correlated with TeV halos~\citep{2012arXiv1202.1455M}. Compact PWN are more accessible to X-ray telescopes due to their larger magnetic fields and higher surface brightnesses. In particular, since the peak synchrotron flux in X-ray halos falls below the Chandra band, the fraction of the total synchrotron emission entering the Chandra band in the X-ray PWN will rise as $\propto$B$^2$. For example, observations of Geminga find a bright PWN with a size of only $\sim$1'~\cite{Posselt:2016lot}. The X-ray flux from this region (0.3-8.0~keV) is $\sim$7.51~$\times$~10$^{-13}$~erg~cm$^{-2}$~s$^{-1}$, which lies far above the Chandra sensitivity threshold for a source of this size. Models of this PWN indicate that the magnetic field strength is $\sim$20~$\mu$G, significantly larger than the average interstellar medium value~\cite{2010ApJ...715...66P, Posselt:2016lot}. If the intensity ratio of the compact X-ray PWN and TeV halo is consistent for Geminga-like pulsars, it is likely that Chandra observations could detect PWN near the center of TeV halos, despite the fact that they would be unable to detect the broader X-ray halo.

While X-ray observations are unable to determine the distance to a PWN, additional radio observations of PWN can potentially provide distance information through HI absorption measures~\citep[e.g.][]{Leahy:2007qw}. Recent studies indicate that n$_H$ measurements can also be correlated with X-ray absorption features in PWN~\citep{He:2013zwx}. In both cases the uncertainties in PWN distance measurements are considerable. However, these observations can potentially indicate which TeV halos are most likely to be located near Earth, and thus require more careful follow-up.

\subsection{Thermal Emission from Misaligned Pulsars}
\label{sec:optical}

While PWN observations offer the ability to conclusively determine the TeV halo origin of HAWC sources, the combination of TeV halo and X-ray PWN observations will be unable to lift the degeneracy between the pulsar distance and its luminosity. For pulsars with beams oriented towards Earth, the dispersion in the radio pulse is typically employed to calculate the pulsar distance~\citep{Cordes:2002wz}. However, for misaligned pulsars, the neutron star point source is likely to be visible only through its thermal emission.

Fortunately, observations of thermal neutron star emission can constrain the distance to the nearest pulsars, due to the correlation between the temperature and luminosity of a blackbody. The thermal evolution of $10^4-10^6$ year old pulsars is an area of active research \cite{Page:2004fy,Page:2009fu}. At birth, pulsars are believed to reach temperatures of $10^{10}$ K and rapidly cool through neutrino emission to temperatures of $T \lesssim 10^{7}$ K within $10^4$ years. In the absence of additional heat sources, after $\sim$10$^4$ years the photon emission of the neutron star can be modeled as a blackbody with a luminosity $L$~=~4$\pi$R$\sigma_{B}T^4$ 
%% While most TeV halos are too distant to produce detectable thermal emission, HAWC observations of the TeV halos in closest proximity to Earth are expected to be particularly fruitful. 
%\begin{equation}
%L = 4 \pi R \sigma_B T^4,
%\end{equation}
where $L$, $R$ and $T$ are the luminosity, radius, and blackbody temperature of the neutron star for a distant observer. For pulsars like Monogem and Geminga, with ages \mbox{(1-3)$\times10^5$~yr}, the blackbody emission depends on the composition of the neutron star. Neutron star crusts composed of heavier elements inhibit cooling at lower temperatures, implying a higher luminosity and temperature for middle-aged pulsars. Depending on these factors, $\sim$10$^5$ year old neutron stars should have $T_{\infty} \simeq 10^{5.2}-10^{6.2}$~K, corresponding to blackbody luminosities $L_{\infty}$~$\simeq$~10$^{30}$ --- 10$^{34}$~{\rm ergs s$^{-1}$}. %While all current observations fall within this range of temperatures and luminosities \cite{Prinz:2015jkd}, it is possible that, through magneto-thermal heating, pulsars with anomalously large initial magnetic fields may be hotter after $10^5$ years. Depending on the model for magnetic heating of the neutron star fluid, a neutron star with an initial magnetic field of $10^{15} G$ could have a blackbody temperature as high as $10^{6.2}$ Kelvin after $10^5$ years.

Detections of neutron star thermal emission have typically been obtained via X-ray observations of young systems. This technique has previously been employed to follow-up the ``Magnificent Seven"  neutron stars first observed by ROSAT~\citep{1997Natur.389..358W, 2001ASPC..234..225T}. For a neutron star with a temperature of 10$^6$~K, the peak in the blackbody spectrum occurs at 243~eV, producing a significant thermal flux in the energy range observable by existing X-ray instrumentation. While neutron stars with temperatures of $\sim$10$^5$~K have blackbody spectral peaks which fall somewhat below the X-ray band, a significant tail of X-ray emission will still be observable.

The sensitivity of Chandra and XMM-Newton to X-ray emission from $10^4-10^5$ year old pulsars will depend on local backgrounds and observing conditions. However, publicly available point-source sensitivity estimates provided by Chandra and XMM \cite{Chandra, XMM}, can be used to determine the detection prospects for neutron stars at various temperatures and distances. While the exact sensitivity of any X-ray observation depends strongly on the local background, in general the ACIS instrument on Chandra can resolve point sources that deposit $4 \times 10^{-15}~{\rm ergs~ cm^{-2}~s^{-1}}$ over $0.4-6$ keV energies after $10^4$ seconds of integration \cite{Chandra}. Using the overall ACIS efficiency for accepting photons (Figure 6.3 of \cite{Chandra}), and convolving this with the high energy tail of the blackbody spectrum of a $\sim10^6$ K neutron star, Chandra could find neutron stars out to $\sim 3$ kpc in $10^4$ seconds. Because these measurements are quite sensitive to the fraction of the neutron star black body spectrum that exceeds the minimum energy cutoff, this sensitivity drops off sharply for lower temperature neutron stars. For example, Chandra would find $5 \times 10^5$ Kelvin neutron stars only out to $200$~pc in $10^4$ seconds. Similarly, the XMM-Newton EPIC instrument typically obtains $5 \sigma$ sensitivity to sources with an integrated flux of $3 \times 10^{-15}~{\rm ergs~ cm^{-2}~s^{-1}}$ in the energy range $0.5-2$ keV for a $10^4$ second integration time \cite{XMM}.

In addition to X-ray observations of thermal neutron star emission, we also consider the potential for optical observations to constrain the distance to nearby pulsars. While optical telescopes are typically less sensitive to thermal neutron star emission, we consider this avenue for two reasons. First, a number of high sensitivity optical telescopes will come online over the next few years. Second, optical observations have higher angular resolution, potentially providing parallax distances for nearby pulsars.

The direct optical observation of a neutron star candidate inside a TeV halo is challenging, due to the large ($\sim$2$^\circ$) region of interest that optical surveys must cover. These observations would require survey instrumentation. Among the most sensitive current datasets is the DECam Legacy survey data, which overlaps a sizable portion of the HAWC field of view. A T~=~$10^6$~K neutron star has an absolute g-band magnitude of 19.8, assuming a representative neutron star size of $R_\infty~=~10$~km. Assuming a limited survey magnitude of 25.0, this only enables us to observe neutron stars out to a distance of $\sim$100~pc. It is relatively improbable that such a neutron star exists. However, as data from these catalogs is already available, it will be easy to quickly survey a large ensemble of TeV halos looking for extremely local pulsar associations.

Fortunately, deeper observations are possible in targeted searches. In this case the putative pulsar must first be localized based on the observation of an X-ray PWN, an observation of thermal X-ray emission, or through a careful study of the correlation between the location of observed pulsars and TeV halo morphologies. Using Hubble Space Telescope observations, limiting magnitudes between 28 and 29 are possible, depending on the observation duration and survey mode. These observations would allow for optical detection of neutron stars out to distances of $\sim$350---500~pc. Given the high angular resolution of optical measurements, this would still  allow for accurate distance measurements via parallax, offering a unique chance to identify hidden pulsars in close proximity to Earth.

%\begin{figure*}[tbp]
%\centering
%\includegraphics[width=.48\textwidth]{f2_ssc_flat.pdf}
%\includegraphics[width=.48\textwidth]{f2_ssc_flat.pdf}
%\caption{Figure Text} 
%\label{fig:variance}
%\end{figure*}

%\section{Discussion}
%\label{sec:discussion1}

%\subsection{Middle-Aged TeV Halos}
%\label{sec:oldpwn}
%Throughout the majority of this paper we have concentrated our analysis on middle-aged pulsars, which are uniquely accessible to HAWC instrumentation. In this section, we additionally point out several advantages in investigations aimed at these systems, compared to systems with higher spin-down powers. First, we note that the morphological differences between the 

\section{Young TeV Halos}
\label{sec:youngpwn}

Throughout Sections~\ref{sec:hawcsensitivity}~through~\ref{sec:pwn}, we have concentrated only on middle-aged pulsars. However, of the 15 TeV halo candidates listed in Tables~\ref{tab:HAWCsources}~and~\ref{tab:HAWCsourcesyoung}, 10 have ages below the 100~kyr cutoff used in this study, and 5 have ages below 20~kyr. Moreover, H.E.S.S. observations have detected 19 TeV halos (and 20 additional potential TeV halos), 19 (15) of which are associated with pulsars younger than 100~kyr. This is not unexpected, as these sources have the largest spin-down powers 

For the moment, we concentrate on HAWC sources and optimistically assume that each of these of these associations is real. Taking the beaming fractions for each of these 10 systems, as calculated in Equation~\ref{eq:beaming}, we predict that an additional 23$^{+13}_{-9}$ mis-aligned, TeV halos with ages below 100~kyr should already be observable by HAWC. Indeed, there is evidence that at least one such system, 2HWC J1955+285, does exist in the HAWC data, owing to $\gamma$-ray observations that confirm a pulsar origin. Projecting these observations to 10 years of HAWC data indicate a possible contribution of 194$^{+26}_{-23}$ TeV halos. 

These numbers almost certainly overestimate the contribution of TeV halos to this population of young sources.  If all 15 young and middle-aged TeV halo candidates listed in Tables~\ref{tab:HAWCsources}~and~\ref{tab:HAWCsourcesyoung} had emission dominated by TeV halo activity, we would anticipate a current population of 54$^{+15}_{-13}$ TeV halos in the 2HWC catalog. This exceeds the total population of 39 2HWC sources, and indicates that some of the associations in Tables~\ref{tab:HAWCsources}~and~\ref{tab:HAWCsourcesyoung} are likely due to a convolving factor, such as an SNR that emits isotropically. While we argued in Section~\ref{sec:tevhalos} that these sources are distinct, at young ages they may have similar radial extents and be difficult to differentiate in TeV observations. A similar result is found for TeVCAT sources observed by both H.E.S.S. and VERITAS. At present, 30 sources coincident with known PWN are listed in these catalogs. Since the majority of these sources are young, a beaming fraction of $\sim$30\% is appropriate. This would predict an underlying population of 70$^{+18}_{-16}$ unidentified TeV halos would be expected to be observed in the H.E.S.S. and VERITAS data. However, only 33 such sources are observed. We stress that while this comparison is illuminating, it should not be taken at face value, as VERITAS and H.E.S.S. are pointed instruments, and have made deeper observations of regions with associated pulsar sources.

However, the population of young TeV halos is particularly intriguing. The study of these systems provides a unique handle constraining both the spin-down evolution of young pulsars and the e$^+$e$^-$ injection spectrum in young systems. In particular, Equation~\ref{eq:losstime} shows that the energy loss time of  e$^+$e$^-$ varies inversely with energy. Since the inverse-Compton scattering of $\gtrsim$10~TeV e$^+$e$^-$ occurs primarily near the Klein-Nishina limit, the observed $\gamma$-ray spectrum is an proxy for the e$^+$e$^-$ energy. The $\gamma$-ray spectrum encodes the e$^+$e$^-$ spectrum. However, in young systems the spin-down power of the pulsars evolves significantly within a the cooling time of e$^+$e$^-$. The steady-state e$^+$e$^-$ spectrum of young TeV halos is thus dependent on the braking index of young pulsars. In particular, younger pulsars with higher braking indexes will be expected to have softer TeV halo spectra.

%Additionally, the spectral evolution of TeV halos potentially encode important information concerning both the e$^+$e$^-$ injection spectrum and spin-down evolution of middle-aged pulsars. As shown in Equation~\ref{eq:losstime}, the energy loss time of e$^+$e$^-$ varies inversely with energy. Because the inverse-Compton scattering of these particles occurs near the Klein-Nishina regime, the $\gamma$-ray energy is approximately equal to the e$^+$e$^-$ energy, and the $\gamma$-ray spectrum thus implies the relevant cooling times of e$^+$e$^-$ as a function of energy. For the case of relatively old pulsars, such as Geminga, this is a small correction, as the pulsar is unlikely to evolve significantly on the $\sim$20~kyr timescales probed by the energy loss. However, for young pulsars, the change in spin-down luminosity over a cooling timescale is likely to be significant. This allows for TeV halo observations to test models for the braking index and the evolution of the e$^+$e$^-$ injection from young pulsars. 

%However, for many young pulsars, the Geminga-like model predicts significant TeV halo emission, and thus the e$^+$e$^-$ efficiency and spin-down evolution of these systems should be carefully considered. Indeed, it is possible that one such system, 2HWC J1955+285, already exists in the HAWC data. 

\section{Conclusions}
\label{sec:discussion}

Observations by H.E.S.S. have indicated that the luminosity of TeV halos is correlated to the spin-down power of the pulsar generating them. HAWC observations indicate that this trend continues for pulsars with significantly smaller spin-down luminosities. By employing such a model, and utilizing Geminga as a standard candle, we have demonstrated that HAWC observations open a vast new parameter space for TeV halo detection. In particular, HAWC is likely to detect 37$^{+17}_{-13}$ middle-aged TeV halos that are not currently associated with any presently known radio pulsar. Very few of these systems are likely to have been previously detected as $\gamma$-ray pulsars or X-ray PWN. Additionally, HAWC may detect as many as $\sim$100~TeV halos corresponding to pulsars with ages below $\sim$100~kyr that are not associated with any known radio pulsar. The expected number of young TeV halos depends sensitively on the degree of source confusion between the SNR and TeV halo in young systems. Follow up observations will be necessary to determine the nature and proximity of these sources. %We note that the population of TeV halos described here may significantly underestimate the true population, as we have conservatively assumed that pulsars younger than Monogem or older than Geminga do not produce bright TeV halo activity, while HESS observations indicate that a significant population of young TeV halos exists and will also be detectable by HAWC observations. 

%Implications for studies of the Pulsar beams
\subsection{Constraints on Pulsar Evolution}
One significant advantage of pulsars originally discovered as TeV halos is the relatively unbiased detection probability offered by HAWC's wide field-of-view and the isotropic emission of TeV halos. The detection of such systems will allow for improved observational constraints on the size of pulsar beams at both radio and  $\gamma$-ray energies. The angular extent of the $\gamma$-ray beam for young pulsars, in turn, has significant implications for our understanding of the e$^+$e$^-$ acceleration regions in young pulsars~\citep{Gonthier:2003qr, Pierbattista:2014ona, Pierbattista:2016myq}.

Additionally, because TeV halos are expected to be powered by the spin-down luminosity of the associated pulsar, the evolution of the TeV halo luminosity provides information concerning the evolution of a pulsars e$^+$e$^-$ conversion efficiency and the pulsar braking index. As discussed in Section~\ref{sec:youngpwn}, these constraints will be particularly useful for understanding the young pulsar population, since young pulsars are believed to evolve considerably over the cooling timescale of $\sim$TeV e$^+$e$^-$.

%Implications for the Interpretation of the positron excess

\subsection{Cosmic-Ray Diffusion in TeV Halos}
The null observations of TeV halos coincident with the pulsars B1951+32 and J1740+1000 is slightly unsettling, as it indicates dispersion in the correlation between the pulsar spindown luminosity and the observed luminosity of the TeV halo. H.E.S.S. obtains similar results, recording null observations of TeV halos surrounding five ATNF pulsars with spin-down luminosities exceeding $\sim$10$^{36}$~erg~s$^{-1}$, and distances smaller than $\sim$10~kpc~\cite{Abdalla:2017vci}. Our Geminga-like model would predict that these systems are detectable by current observations. 

This dispersion may be introduced by two separate mechanisms, both of which are theoretically interesting. The first, advocated by the H.E.S.S. collaboration~\cite{Abdalla:2017vci}, is that the conversion efficiency of spin-down power to e$^+$e$^-$ injection may vary considerably between different pulsars. This may, in turn, have significant implications for our understanding of both pulsar acceleration models, and our understanding of the pulsar contribution to the cosmic-ray positron excess. 

The second possibility is that the diffusion environment may vary significantly within the TeV halos surrounding different pulsars.  As discussed in Section~\ref{sec:tevhalos}, if a pulsar does not produce a region with significantly inhibited diffusion (compared to the surrounding interstellar medium), the angular scale of TeV halos (compared to current observations) would be stretched by more than two orders of magnitude.  These ultra-diffuse TeV halos would be inaccessible to ACT observations, and would be difficult to detect with HAWC, due to their extremely low surface brightnesses. There is some evidence for this second scenario. In particular, variations in the size of and properties of TeV halos can be motivated motivated by the dependence of both SNR and PWN morphologies on the surrounding interstellar medium.

The effect of supernovae, and their pulsar progeny, on the local diffusion of cosmic rays is of significant interest. H.E.S.S. observations of TeV halos indicate that the size of the TeV halo grows as a function of time. However, spatial regions with inhibited diffusion can not persist for periods $\sim$1~Myr over regions of $\sim$100~pc without significantly affecting the average cosmic-ray diffusion parameters of the Milky Way~\cite{2017arXiv170208436H, Trotta:2010mx}. Thus, the impact of TeV halos on the efficiency of cosmic-ray diffusion is squeezed by observational constraints from both directions. The HAWC detection of TeV halos surrounding Geminga and Monogem are of particular importance, as these systems are significantly longer-lived than previously known TeV halos. Future observations of TeV halos will allow models to correlate the observable parameters of each system with the resulting TeV halo size. These measurements will have important implications for our understanding of cosmic-ray propagation within, and in close proximity to, TeV halos.

% This dispersion may be introduced by two separate mechanisms, both of which are theoretically interesting. The first is that the efficiency of the conversion from pulsar spindown power to e$^+$e$^-$ injection varies significantly between pulsars, a result which would have significant implications for our understanding of the pulsar contribution to the cosmic-ray positron excess. The second is that the diffusion environment varies significantly in the regions surrounding TeV halos, producing TeV halos of different sizes and surface brightnesses. In particular, if some pulsars do not produce surrounding regions with significantly inhibited cosmic-ray propagation, the TeV halos from these sources would have angular extents more than an order of magnitude larger than expected. These may be difficult to detect even with wide field-of-view instruments such as HAWC. 

\subsection{The Cosmic-Ray Positron Excess}
Finally, we note that observations, or null observations, of nearby mature pulsars can significantly affect our interpretation of the cosmic-ray positron excess observed by PAMELA and AMS-02~\citep{Adriani:2008zr, Aguilar:2013qda}. In the past, dark matter~\citep{Cirelli:2008pk, Cholis:2008wq, ArkaniHamed:2008qn, Bergstrom:2008gr}, pulsar~\citep{Hooper:2008kg, Profumo:2008ms, Malyshev:2009tw, Linden:2013mqa, Cholis:2013psa}, and stochastic acceleration~\citep{2009PhRvL.103e1104B, Mertsch:2009ph, Ahlers:2009ae, Cholis:2013lwa, 2011ApJ...733..119K, Cholis:2017qlb} models have all provided plausible fits to the observed data. However, the recent observations of TeV halos have indicated, for the first time, that pulsars are guaranteed to provide the necessary e$^+$e$^-$ flux to drive the rising cosmic-ray positron fraction, and that low-energy e$^+$e$^-$ are likely to escape into the interstellar medium~\citep{2017arXiv170208436H}. 

The unbiased observation of nearby pulsars is critical to predict the high-energy behavior of the rising cosmic-ray positron fraction, as the e$^+$e$^-$ flux above $\sim$500~GeV is dominated by the nearest middle-aged sources. In particular, the cutoff energy of the pulsar contribution to the e$^+$e$^-$ spectrum can be calculated based on the age, proximity and spin-down luminosity of the nearest pulsars. Recently, Fermi-LAT observations have indicated that the hardening of the combined e$^+$e$^-$ spectrum continues until at least 2~TeV~\citep{Abdollahi:2017kyf}, making a complete understanding of the nearby pulsar population even more imperative. Additionally, if a small number of pulsars produce the highest energy leptons, then ``wiggles" may be observable in the cosmic-ray positron fraction and total lepton spectrum~\citep{Profumo:2008ms, Shaviv:2009bu, Malyshev:2009tw}, an observation which could definitively differentiate models of the cosmic-ray positron excess. However, these features can \emph{only} be correlated with the positions and ages of known pulsars in the case that the entire nearby pulsar population is known. The observation of the closest TeV halos thus play an important role in producing smoking gun evidence in favor of a pulsar origin of the cosmic-ray positron excess.

\section*{Acknowledgements}
We thank Brenda Dingus, Marc Kamionkowski, Anna Nierenberg, Paul Ray,  Joe Silk, Pat Slane, Hao Zhou, and Haochen	Zhang for helpful comments which substantially improved the quality of this manuscript. We especially thank John Beacom for numerous comments and suggestions concerning the interpretation of these results.
TL acknowledges support from NSF Grant PHY-1404311 to John Beacom. IC is supported by NASA Grant NNX15AB18G and the Simons Foundation. Research at Perimeter Institute is supported by the Government of Canada through Industry Canada and by the Province of Ontario through the Ministry of Economic Development \& Innovation.

\bibliography{hawc_ns}

\end{document}